\documentclass[prb,twocolumn,showpacs,floatfix,amsmath,amssymb,superscriptaddress]{revtex4-2}
\usepackage{amsfonts}
\usepackage{stmaryrd}
\usepackage{bbm}
\usepackage{mathrsfs}
\usepackage{tipa}
\usepackage{amssymb}
\usepackage{txfonts}
\usepackage{graphicx}
\usepackage{dcolumn}
\usepackage{epstopdf}
\usepackage[colorlinks,linkcolor=blue,urlcolor=blue,citecolor=blue]{hyperref}
\usepackage{xcolor}
\usepackage{multirow}
\usepackage{subfigure}
\usepackage{url}
\usepackage{upgreek}
\usepackage[utf8]{inputenc}
\usepackage[english]{babel}
\usepackage{siunitx}
\usepackage[version=4]{mhchem}
\usepackage{bm}

\begin{document}

\newcommand*{\icm}{cm$^{-1}$}
\newcommand*{\Tc}{T$_c$}

\title{Magnetization processes and spin dynamics across field-induced phase transitions in the quasi-two-dimensional quantum magnet Cu$_2$(OH)$_3$Br}

\author{Anneke~Reinold}
\affiliation{Department of Physics, TU Dortmund University, 44227 Dortmund, Germany}

\author{Dirk~Wulferding}
\affiliation{Department of Physics and Astronomy, Sejong University, Seoul 05006, Republic of Korea}

\author{Laur~Peedu}
\author{Kirill~Amelin}
\author{Urmas~Nagel}
\author{Toomas~Rõõm}
\affiliation{National Institute of Chemical Physics and Biophysics, 12618 Tallinn, Estonia}

\author{Zhiying~Zhao}
\affiliation{State Key Laboratory of Structural Chemistry, Fujian Institute of Research on the Structure of Matter,
Chinese Academy of Sciences, Fuzhou, Fujian 350002, China}

\author{Patrick~Pilch}
\author{Changqing~Zhu}
\affiliation{Department of Physics, TU Dortmund University, 44227 Dortmund, Germany}

\author{Hans~Engelkamp}
\affiliation{HFML-FELIX, Toernooiveld 7, 6525 ED Nijmegen, the Netherlands}

\author{Lucas Berger}
\affiliation{Institute of Physics II, University of Cologne, D-50937 Cologne, Germany}

\author{Denis~I.~Gorbunov}
\author{Yurii~Skourski}
\affiliation{Hochfeld-Magnetlabor Dresden (HLD-EMFL), Helmholtz-Zentrum Dresden-Rossendorf, 01328 Dresden, Germany}

\author{Kwang-Yong Choi}
\affiliation{Department of Physics, Sungkyunkwan University, Suwon 16419, Republic of Korea}

\author{Thomas~Lorenz}
\affiliation{Institute of Physics II, University of Cologne, D-50937 Cologne, Germany}

\author{Zhe~Wang}
\affiliation{Department of Physics, TU Dortmund University, 44227 Dortmund, Germany}

\date{\today}

\begin{abstract}

We present magnetic field-dependent evolution of magnetization and spin dynamics in the quasi-two-dimensional spin-$1/2$ magnet \ce{Cu2(OH)3Br}, consisting of alternately coupled ferromagnetic Cu1 and antiferromagnetic Cu2 spin chains.
Terahertz spectroscopy reveals a pronounced field-direction dependence of the low-energy magnetic excitation spectrum.
For magnetic fields applied perpendicular to the spin chains, $B\parallel a$ and $B\parallel c^*$, the spectra undergo abrupt reconstructions at the spin-flop transitions identified independently by high-field magnetization measurements.
For $B\parallel b$, by contrast, no spin-flop occurs; instead, the excitation spectrum evolves continuously with field and exhibits a strong terahertz radiation polarization dependence as the ferromagnetic Cu1 subsystem becomes progressively polarized.
At higher fields, the complex low-field spectrum is replaced by a reduced set of broad excitations, consistent with a weakening of the coupling between the ferromagnetic and antiferromagnetic chain subsystems.
Complementary Raman spectroscopy resolves magnon and spinon excitations alongside several phonon modes and traces the characteristic temperature and magnetic-field dependent evolution of the magnetic excitations.
The combined spectroscopic and magnetization results map out how a magnetic field reorganizes the coupled ferromagnetic and antiferromagnetic subsystems in \ce{Cu2(OH)3Br} across field-induced phase transitions.

\end{abstract}

\maketitle

\section{\label{sec:Introduction}Introduction}

\typeout{Text width: \the\textwidth}
\typeout{Column width: \the\columnwidth}
\typeout{Column separation: \the\columnsep}

In low-dimensional quantum magnets, reduced dimensionality and strong quantum fluctuations can fundamentally alter the nature of elementary excitations. 
While conventional magnetically ordered systems are typically described by collective spin-wave excitations carrying spin~$S=1$, one-dimensional quantum spin systems can instead host fractionalized quasiparticles known as spinons carrying spin~${S=1/2}$~\cite{Faddeev_1981,Mourigal_2013}.
In antiferromagnetic spin-${1/2}$ chains, these excitations often manifest as broad excitation continua rather than well-defined sharp dispersive modes~\cite{Mourigal_2013,Lake_2005}.
Understanding the formation, stability, and interactions of such fractionalized excitations is a central challenge in quantum magnetism and strongly correlated electron systems~\cite{Lake_2010,WangDeisenhofer15,WangLoidl17}.
Particularly intriguing are systems in which fractionalized and conventional collective excitations coexist. Interactions between different quasiparticles can give rise to emergent collective phenomena and unconventional excitation spectra~\cite{Coldea_2010,WangLoidl2018,Faure2018,Zhang20,Sahasrabudhe20,
Testa_2021,Zhang_2020,WangKollath2024,Choi_2026}.

\begin{figure*}[t]
\centering
\includegraphics[width=1\linewidth]{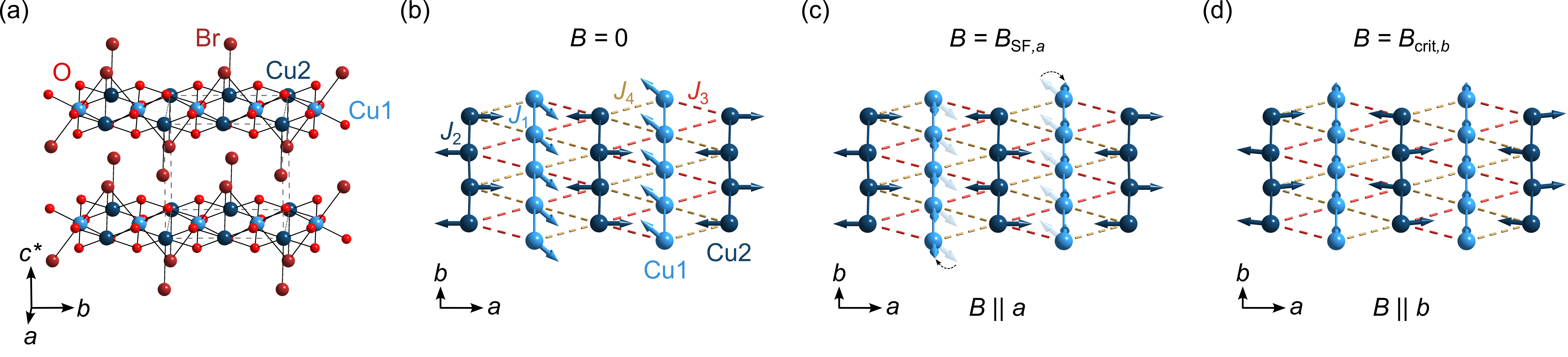}
    \caption{(a) Layered crystal structure of \ce{Cu2(OH)3Br}. The two inequivalent copper sites Cu1 and Cu2 are shown in light and dark blue balls, respectively, while oxygen O and bromine Br atoms are displayed in red and brown. Hydrogen atoms are omitted for clarity.
    (b) Schematic magnetic structure in the magnetically ordered zero-field phase. 
    (c) Schematic illustration of the spin-flop phase for a magnetic field applied along the $a$ axis. In the spin-flop phase, the Cu1 moments are oriented approximately along the $b$ axis, with a slight canting toward the applied magnetic field, while the orientation of the Cu2 moments remains largely unchanged.
    (d) Schematic high-field magnetic structure for $B\parallel b$. The Cu1 moments are fully polarized along the applied magnetic field, while the Cu2 moments remain predominantly aligned along the $a$ axis and are only slightly canted toward the field direction.}
    \label{fig:crystalstructure}
\end{figure*}

Low-dimensional quantum magnets combining ferromagnetic and antiferromagnetic correlations therefore provide a promising platform for investigating the interplay between fractionalized and collective magnetic excitations.
The botallackite compound \ce{Cu2(OH)3Br} has recently been identified as a realization of such a system.
\ce{Cu2(OH)3Br} crystallizes in the monoclinic space group $P2_1/m$ and consists of two crystallographically inequivalent spin-$1/2$ \ce{Cu^{2+}} sites, denoted Cu1 and Cu2, which form alternating chains along the crystallographic $b$ direction [see Fig.~\ref{fig:crystalstructure}(a)(b)] \cite{Oswald_1961, Aebi_1948}.
The magnetic interactions are highly anisotropic: while the Cu2 chains are governed by a strong antiferromagnetic exchange interaction, the Cu1 chains exhibit weaker ferromagnetic coupling.
Additional couplings between neighboring chains connect these magnetic building blocks into a distorted triangular network within the $ab$ plane, introducing frustration and giving rise to a quasi-two-dimensional magnetic system.
Below $T_\mathrm{N}=\SI{9.3}{K}$ a long-range magnetic order is formed, which is characterized by ferromagnetically aligned Cu1 spins tilted by approximately $45^\circ$ from the $a$ axis within the $ac^*$ plane, whereas the spins in the Cu2 chains are aligned antiferromagnetically [see Fig.~\ref{fig:crystalstructure}(b)].
Experimental estimates indicate a pronounced spatial anisotropy of the magnetic exchange interactions in botallackite, with the corresponding exchange paths $J_{1,2,3,4}$ indicated in Fig.~\ref{fig:crystalstructure}(b).
Recent quantum Monte Carlo simulations of the magnetization \cite{Reinold_2025}  and inelastic neutron scattering data \cite{Zhang_2020} determine the dominant antiferromagnetic exchange along the Cu2 chains $J_2=\SI{5.3}{\milli\electronvolt}$, the ferromagnetic exchange in Cu1 chains $J_1=-0.3J_2\approx\SI{-1.6}{\milli\electronvolt}$, and the dominant interchain coupling $J_3=0.2J_2\approx\SI{1.1}{\milli\electronvolt}$, while the interchain exchange $J_4$ is negligibly small~\cite{Reinold_2025,Khatua_arXiv_2026}.
Despite the formation of three-dimensional long-range magnetic order~\cite{Zheng_2009, Zhao_2019, Zhang_2020, Xiao_2022}, pronounced low-dimensional magnetic correlations persist~\cite{Zhang_2020, Reinold_2025, Khatua_arXiv_2026}.

Inelastic neutron-scattering measurements have revealed the coexistence of a spinon continuum associated with the antiferromagnetic Cu2 chains and magnon excitations originating from the ferromagnetic Cu1 chains, together with signatures of spinon-magnon interactions mediated by finite interchain coupling~\cite{Zhang_2020}.
High-field magnetization measurements have determined field-induced phase transitions for different field orientations and found a half-magnetization plateau \cite{Zhao_2019,Xiao_2022,Povarov_2024,Reinold_2025}.
High-field electron-spin-resonance (ESR) measurements for frequencies up to 900~GHz ($\simeq 3.7$~meV) and applied magnetic fields up to 16~T have mapped out low-energy magnetic excitations in different phases~\cite{Xiao_2022,Povarov_2024}.

In particular, magnetic field applied along the crystallographic $b$ axis polarizes the spins in the ferromagnetic chains above $B_{\mathrm{crit},b} = \SI{16.3}{T}$, while the spins in the antiferromagnetic chains remain predominantly perpendicular to the field, leading to an effective decoupling of the two chain subsystems and a concomitant dimensional reduction~\cite{Reinold_2025}.
The magnetization along the $b$ axis exhibits an extended plateau-like regime near one-half of the saturation magnetization, corresponding to the field-induced polarization of the ferromagnetic chain subsystem~\cite{Reinold_2025}.
These observations highlight the coexistence of multiple energy scales and competing magnetic interactions in \ce{Cu2(OH)3Br}, establishing it as a rare model system for studying coupled quasiparticle dynamics in a low-dimensional quantum magnet~\cite{Zhang_2020,Khatua_arXiv_2026}.

While inelastic neutron scattering directly probes the magnetic excitation spectra for finite momentum- and energy-transfers, optical spectroscopies provide complementary polarization-resolved access to low-energy excitations at near-zero momentum transfer.
In this work, we use terahertz and Raman spectroscopies to investigate long-wavelength collective modes and magnetic continua for different polarizations, applied field orientations, and temperatures above and below $T_N$. In particular, the excitations are studied in applied fields up to 31~T and in spectral range up to 90~meV, which provide a more comprehensive picture of the dynamical properties crossing phase transitions. 
The paper is organized as follows: 
After providing the experimental details in Sec.~\ref{sec:ExperimentalDetails}, we present our own magnetization data for different field orientations and temperatures in Sec.~\ref{subsec:Magnetization}.
In Sec.~\ref{subsec:FTIR}, we focus on the broadband terahertz spectroscopic measurements at zero field and high magnetic fields up to 31~T for different polarizations and temperatures. In comparison with the previous ESR measurements, our spectroscopic measurements not only extend into higher fields, but also map out higher energy excitations up to about 7~meV and reveal their polarization dependence.
In Sec.~\ref{subsec:Raman}, we show the Raman spectroscopic data for different temperatures and polarizations and in an applied magnetic field. The Raman spectroscopy covers a broader spectral range up to 90~meV and reveals magnon, spinon, as well as phonon excitations.
The spectroscopic data are discussed in comparison with each other and also with the magnetization data as well as previous inelastic neutron scattering results. The key findings are summarized in Sec.~\ref{sec:Conclusion}.

\section{\label{sec:ExperimentalDetails}Experimental details}

Single crystals of \ce{Cu2(OH)3Br} were grown using the hydrothermal synthesis procedure described in Ref.~\cite{Zhao_2019}.
The crystals had typical dimensions of approximately \mbox{$2 \times 5 \times 0.5$ mm$^3$} for $a \times b \times c^*$, where $c^*$ is perpendicular to the \textit{ab} sample plane.
Magnetization was measured down to 1.45 K using static fields up to \SI{14}{T} and pulsed fields up to \SI{45}{T}.

Terahertz transmission measurements as a function of temperature and magnetic field were performed using a Martin--Puplett interferometer coupled to a ${}^3$He cooled \ce{Si} bolometer.
Different optical filters were used to optimize the bolometer response over the investigated spectral range.
The sample was mounted in a cryostat with a superconducting magnet providing magnetic fields up to \SI{17}{T} and temperatures down to \SI{3}{K}.
The magnetic field was applied along the crystallographic $a$, $b$, or $c^*$ axis [see Fig.~\ref{fig:crystalstructure}(a)], while the terahertz wave vector was along $c^*$ $(\mathbf{k}\parallel c^*)$.
A rotatable wire-grid polarizer determined the direction of linear polarization within the $ab$ plaqne.
Additionally, unpolarized high-magnetic field absorption measurements up to \SI{31}{T} at a temperature of \SI{2}{K} were performed using a Fourier-transform spectrometer (Bruker VERTEX 80V).

Field- and temperature-dependent Raman spectroscopic experiments were performed in a backscattering configuration with a 515 nm laser (Cobolt Fandango 05-01 series) focused onto the sample using a series of achromatic lenses, generating a spot diameter of less than 100 \si{\micro\metre} with a laser power below \SI{0.5}{mW}.
The sample was mounted via silver paint (Ted Pella, Inc.) onto a cold finger inside a closed-cycle magneto-optical cryostat (Oxford SpectroMagPT) with a base temperature of \SI{1.6}{K} and a maximum magnetic field of \SI{7}{T}.
Measurements were carried out with light polarization lying within in the crystallographic ab plane, and with fields applied along the $c^*$ axis.
Polarization control was achieved using superachromatic $\lambda/2$ and $\lambda/4$ waveplates (Thorlabs) for linearly and circularly polarized light configurations, respectively.
Inelastically scattered light was guided into a Czerny-Turner-type spectrometer (Princeton Instruments TriVista 777) and recorded through a charge-coupled device detector (PyLoN eXcelon 100BR).
For additional in-plane polarization-dependent Raman measurements at $T=\SI{5}{K}$ and at $T=\SI{300}{K}$ we used a 561 nm laser (Oxxius LCX-561S).

\section{\label{sec:ExperimentalResults}Experimental results}

\subsection{\label{subsec:Magnetization}Magnetization}

\begin{figure}[t]
    \centering
    \includegraphics[width=1\linewidth]{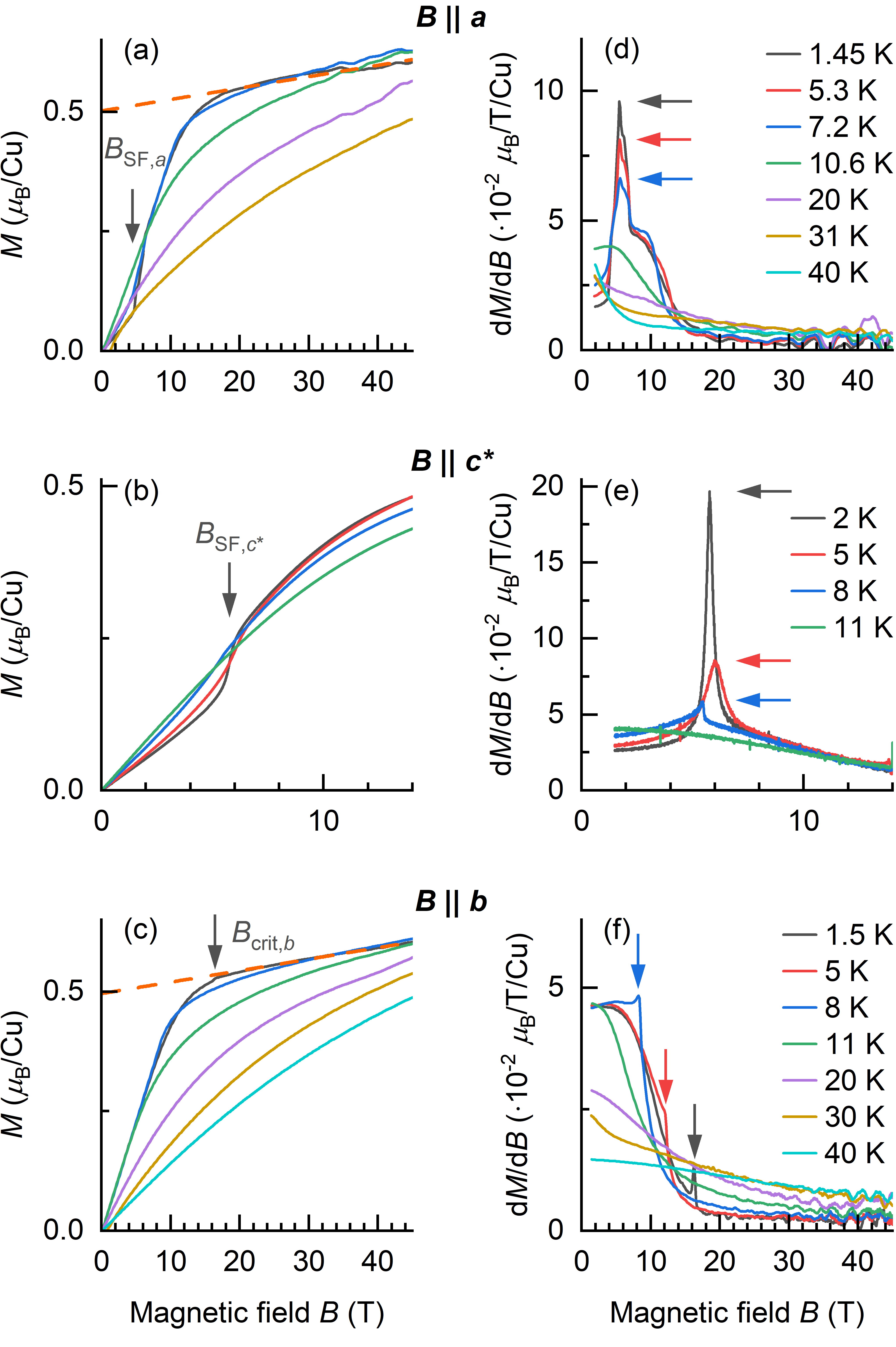}
    \caption{Field-dependent magnetization $M$ and differential magnetic susceptibility $\mathrm{d}M/\mathrm{d}B$ of \ce{Cu2(OH)3Br} measured at different temperatures for different applied magnetic field orientations (a)(d) $B \parallel a$,  (b)(e) $B \parallel c^*$ and (c)(f) $B \parallel b$.}
    \label{fig:Magnetization_abc}
\end{figure}

To establish the field-induced magnetic phases of \ce{Cu2(OH)3Br} as a basis for the subsequent spectroscopic analysis, we first examine its magnetization for different field orientations.
Figures~\ref{fig:Magnetization_abc}(a)--(c) show the magnetization $M(B)$ for $B\parallel a$, $B\parallel c^*$, and $B\parallel b$, respectively, while Figs.~\ref{fig:Magnetization_abc}(d)--(f) display the corresponding differential susceptibilities $\mathrm{d}M/\mathrm{d}B$.

For $B\parallel a$, with the magnetic field applied perpendicular to the alternating ferromagnetic Cu1 and antiferromagnetic Cu2 chains, the magnetization measured at the lowest temperature of 1.45~K initially increases approximately linearly with field.
At $B_{\text{SF},a}=\SI{5.4}{T}$, a distinct step-like feature marks a field-induced spin-flop transition [see arrow in Fig.~\ref{fig:Magnetization_abc}(a)], which gives rise to a pronounced maximum in $\mathrm{d}M/\mathrm{d}B$ in Fig.~\ref{fig:Magnetization_abc}(d).
The feature is observed only in the magnetically ordered phase below $T_\mathrm{N} = \SI{9.3}{K}$~\cite{Zheng_2009, Zhao_2019, Zhang_2020, Xiao_2022}, while its position remains essentially temperature independent.

In the zero-field phase, the ferromagnetically coupled Cu1 moments are oriented approximately along a diagonal of the $ac^*$ plane~\cite{Zhao_2019,Xiao_2022}.
For $B\parallel a$, the spin-flop transition primarily consists of a reorientation of these moments into the field-induced configuration illustrated in Fig.~\ref{fig:crystalstructure}(c), consistent with previous magnetization studies~\cite{Zhao_2019,Xiao_2022,Povarov_2024}.
By contrast, the antiferromagnetically coupled Cu2 moments remain approximately aligned along the $\pm a$ direction because their stronger intrachain exchange resists reorientation.
Above $B_{\mathrm{SF},a}$, the magnetization continues to increase with gradually decreasing slope and approaches a broad high-field regime.
A linear extrapolation of the high-field response of the \SI{1.45}{K} curve to zero field yields an intercept close to $\mu_\mathrm{B}/2$ per Cu, as indicated by the dashed line in Fig.~\ref{fig:Magnetization_abc}(a).
This suggests that the Cu1 spins are essentially fully polarized along the field direction above about \SI{20}{T}, with a high-field differential susceptibility of about $3.3\times10^{-3}\,\mu_\mathrm{B}/\mathrm{T}/\mathrm{Cu}$.

A similar behavior is observed for a magnetic field applied along the crystallographic $c^*$ axis, which is another direction perpendicular to the spin chains.
As shown in Fig.~\ref{fig:Magnetization_abc}(b), the lowest-temperature magnetization curve exhibits a step-like anomaly at $B_{\mathrm{SF},c^*}=\SI{5.8}{T}$, accompanied by a pronounced maximum in $\mathrm{d}M/\mathrm{d}B$ [Fig.~\ref{fig:Magnetization_abc}(e)].
Above $B_{\mathrm{SF},c^*}$, the magnetization continues to increase with a reduced but finite slope.
As for $B\parallel a$, the spin-flop transition is observed only within the magnetically ordered phase below $T_\mathrm{N}$ and becomes progressively less pronounced with increasing temperature.
The corresponding reorientation of the Cu1 moments is analogous to that in the spin-flop phase for $B\parallel a$, Fig.~\ref{fig:crystalstructure}(c), except that the moments are slightly canted toward the $c^*$ axis.
High-field half-magnetization regime, similar to $B\parallel a$, was observed in previous high-field studies~\cite{Zhao_2019,Povarov_2024}.

For $B\parallel b$, which is parallel to the spin chain direction, the magnetization process differs from that for the other two field orientations where the field is perpendicular to the spin chains.
As shown in Fig.~\ref{fig:Magnetization_abc}(c), the low-temperature magnetization initially rises steeply from zero field, with a slope that gradually decreases with increasing field.
Unlike for $B\parallel a$ and $B\parallel c^*$, no distinct low-field spin-flop step is observed.
Instead, at the lowest temperature the magnetization exhibits a kink near $B_{\mathrm{crit},b} = \SI{16.3}{T}$, corresponding to a sharp peak in the differential susceptibility curve in Fig.~\ref{fig:Magnetization_abc}(f), as indicated by the arrows.
With increasing temperature, $B_{\mathrm{crit},b}$ shifts continuously toward lower fields, reaching approximately \SI{8.4}{T} at \SI{8}{K}, and is no longer discernible above $T_\mathrm{N}$.
At higher temperatures, the low-field susceptibility and the magnetization values decrease continuously.
Above $B_{\mathrm{crit},b}$, the magnetization enters a broad high-field regime.
The transition is attributed to a field-induced reorientation of the ferromagnetic Cu1-chain moments toward the applied field, as illustrated schematically in Fig.~\ref{fig:crystalstructure}(d)~\cite{Povarov_2024,Reinold_2025}.
As indicated by the dashed line in Fig.~\ref{fig:Magnetization_abc}(c), extrapolation of the high-field magnetization to zero field yields an intercept close to $\mu_\mathrm{B}/2$ per Cu and a nearly field-independent differential susceptibility of about $2.5\times10^{-3}\,\mu_\mathrm{B}/\mathrm{T}/\mathrm{Cu}$.

\subsection{\label{subsec:FTIR}Terahertz spectroscopy}

\subsubsection{\label{subsec:FTIR_Tdep}Temperature dependence at $B=0$}

Terahertz spectroscopy can probe magnetic excitations through coupling to the THz electromagnetic field.
To investigate the temperature dependence of the excitation spectrum in \ce{Cu2(OH)3Br}, we performed temperature-dependent broadband terahertz transmission measurements in zero magnetic field with the terahertz wave vector fixed along the crystallographic $c^*$ direction, \mbox{$\mathbf{k}\parallel c^*$}, and for two orthogonal polarizations of the THz magnetic field, \mbox{$h^\omega \parallel a$} and \mbox{$h^\omega \parallel b$}.

\begin{figure}[t]
    \centering
    \includegraphics[width=1\linewidth]{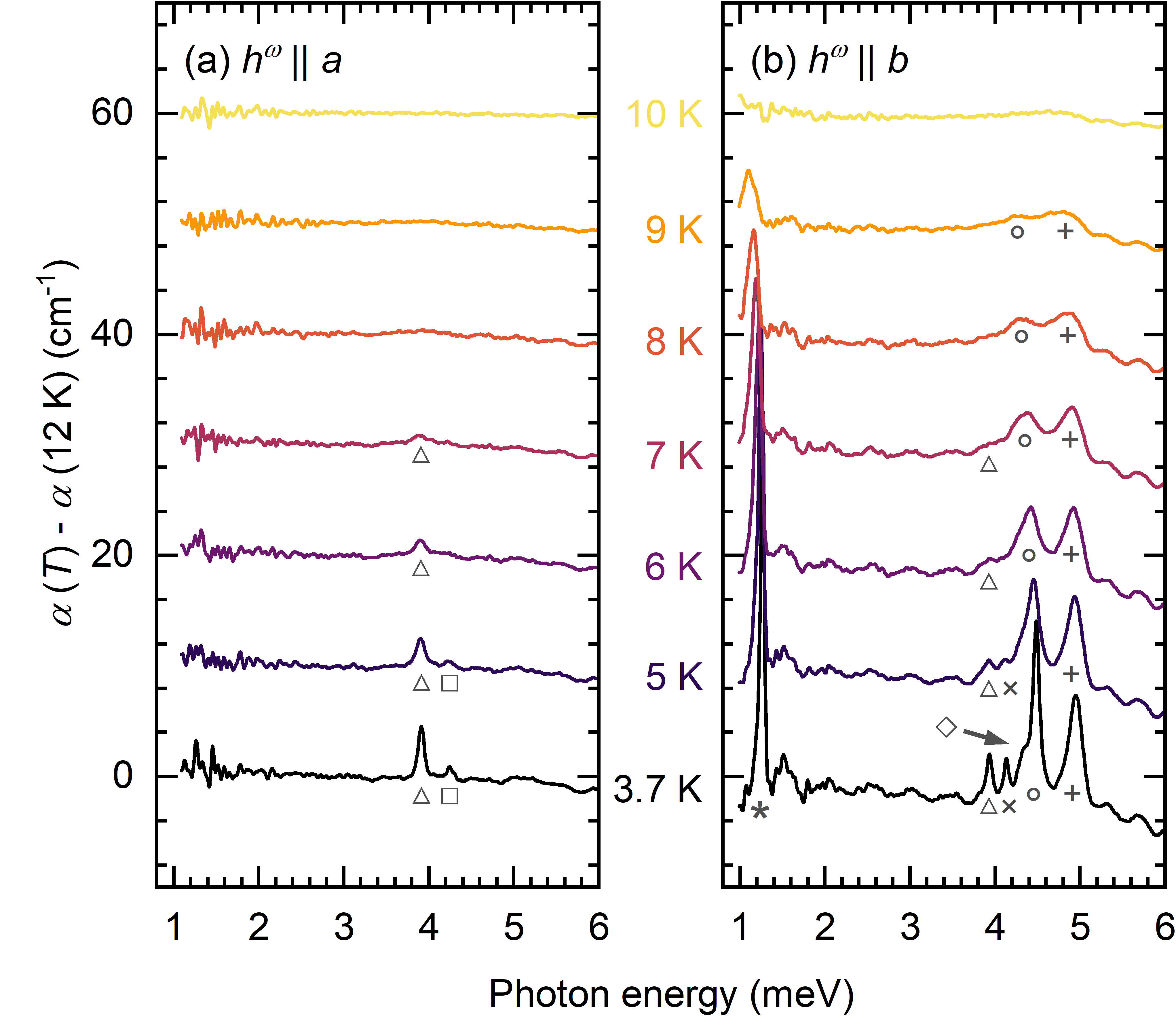}
    \caption{Temperature dependence of the absorption spectrum \mbox{$\alpha(T)-\alpha(\SI{12}{K})$} of \ce{Cu2(OH)3Br} in zero field. (a) and (b) correspond to two orthogonal polarizations \mbox{$h^\omega \parallel a$} and \mbox{$h^\omega \parallel b$} of the incident terahertz radiation. The curves have an offset proportional to the temperature.}
    \label{fig:FTIR_Tdep}
\end{figure}

Figure~\ref{fig:FTIR_Tdep} shows the differential absorption spectra obtained from
\begin{equation}
    \mathrm{\Delta}\alpha(T)=\alpha(T)-\alpha(\SI{12}{K})=-\frac{1}{d}\ln\left(\frac{I(T)}{I(\SI{12}{K})}\right)\,,
    \label{eq:absorptionequation}
\end{equation}
where $d$ is the sample thickness, $I(T)$ is the transmitted intensity measured at temperature $T$, and $I(\SI{12}{K})$ is the reference spectrum measured at $\SI{12}{K}$, i.e. above the Néel temperature $T_\mathrm{N}=\SI{9.3}{K}$.

For \mbox{$h^\omega \parallel a$}, the lowest-temperature spectrum at \SI{3.7}{K} shows a sharp absorption peak near \SI{3.9}{meV}, accompanied by a weaker feature near \SI{4.2}{meV}, marked by the triangle "$\triangle$" and square "$\square$" symbols in Fig.~\ref{fig:FTIR_Tdep}(a), respectively.
Upon warming, both modes broaden and lose intensity.
The 3.9~meV mode remains discernible up to approximately \SI{7}{K}, whereas the weaker 4.2~meV mode is no longer resolvable above about \SI{5}{K}.
Since both excitations are observed only below the magnetic ordering temperature $T_\mathrm{N}=\SI{9.3}{K}$, they are attributed to magnetic excitations of the magnetically ordered phase.
According to inelastic neutron scattering results and simulations of linear spin-wave theory \cite{Zhang_2020}, the 3.9~meV and 4.2~meV modes should correspond to the upper-branch magnon mode and the lower-boundary of spinon excitations, respectively, at the $\Gamma$ point.

In contrast, several absorption peaks emerge for \mbox{$h^\omega \parallel b$} and gain spectral weight upon cooling.
The dominant feature is a sharp and intense mode at \SI{1.3}{meV}, marked by the asterisk "$\ast$" in Fig.~\ref{fig:FTIR_Tdep}(b), which corresponds to a gapped lowest-energy magnon excitation in the ferromagnetic chains as revealed by previous inelastic neutron scattering measurements~\cite{Zhang_2020}.
At higher energies, several modes are observed between approximately \SI{3.9}{meV} and \SI{5.0}{meV}. 
The two weaker modes are located at \SI{3.9}{meV} and \SI{4.1}{meV}, marked by the triangle "$\triangle$" and cross "$\times$" symbols, respectively, whereas two considerably stronger excitations appear at \SI{4.5}{meV} and \SI{4.9}{meV}, indicated by the circle "$\circ$" and plus "$+$" symbols.
The mode marked by the circle exhibits a weak shoulder, indicated by "$\diamond$".
These modes correspond to a cluster of excitations in the energy range where the bottom of the spinon continuum in the antiferromagnetic chains and the top of the magnon band in the ferromagnetic chains overlap with each other, as previously resolved by inelastic neutron scattering measurements \cite{Zhang_2020}.
Upon increasing temperature, all modes gradually lose spectral weight due to enhanced thermal fluctuations.
The low-energy magnon mode as well as the two strong higher-energy excitations remain visible up to approximately \SI{9}{K} and disappear above $T_\mathrm{N}$.
The temperature-dependent behavior confirms the magnetic origin of these modes.

%-------------------------------------------------------------------------------------------------------------------------

\subsubsection{\label{subsec:FTIR_Bdep_a}Field dependence for $B \parallel a$}

\begin{figure}[h!]
    \centering
    \includegraphics[width=1\linewidth]{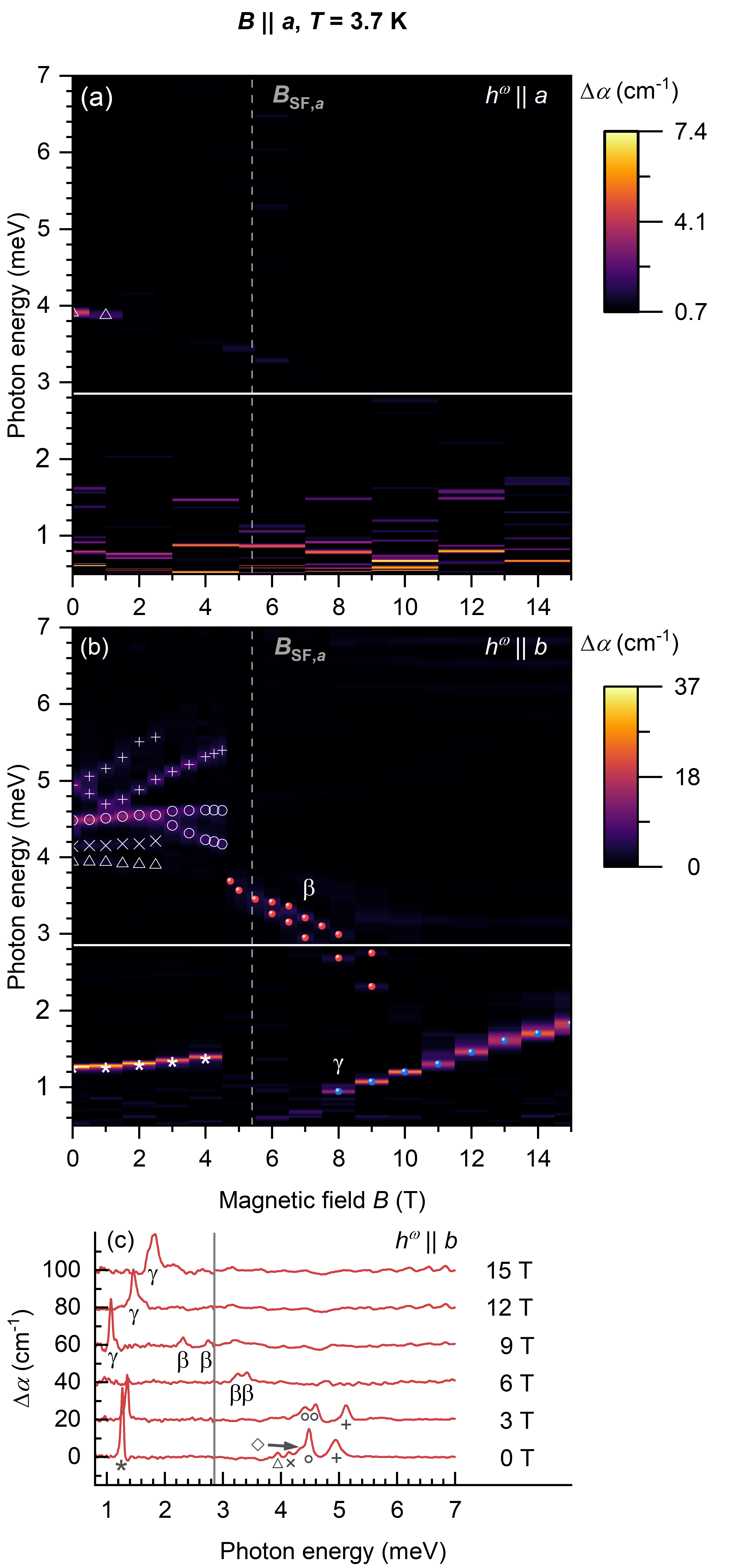}
    \caption{Magnetic-field dependence of the differential absorption spectrum $\Delta\alpha$ of \ce{Cu2(OH)3Br} at \SI{3.7}{K} for $B\parallel a$, where (a) $h^\omega \parallel a$ and (b) $h^\omega \parallel b$.
    The white line marks the boundary between two combined spectral ranges. The spin-flop transition at $B_{\mathrm{SF},a}$ is marked by the grey dashed line.
    (c) Selected spectra for \mbox{$h^\omega \parallel b$} at different magnetic fields. The spectra are offset for clarity.}
    \label{fig:FTIR_Hpa}
\end{figure}

To gain further insight into the nature of the low-energy excitations and their coupling to the magnetic ground state, we investigated the evolution of the absorption spectrum in external magnetic fields applied along three different crystallographic directions.
For all field-dependent measurements, differential absorption spectra were calculated using the corresponding zero-field spectrum as a reference.
The zero field spectrum was recovered by calculating the median of differential absorption  spectra over all  magnetic field values and then the recovered zero field spectrum was added to the differential absorption spectra.
This procedure suppresses field-independent contributions and highlights magnetic-field-induced changes in the excitation spectrum \cite{Zhang20,Amelin20,Amelin2022,Pilch23,Pilch25a,Pilch25b}.

The absorption spectra for $B\parallel a$ are shown in Fig.~\ref{fig:FTIR_Hpa}, and the extracted frequencies of the absorption maxima are summarized in Fig.~\ref{fig:exmodes_maxima}(a).

For \mbox{$h^\omega \parallel a$} [Fig.~\ref{fig:FTIR_Hpa}(a)],
at zero field the excitation near \SI{3.9}{meV} is the only clearly discernible mode, while the square mode observed in Fig.~\ref{fig:FTIR_Tdep}(a) is too weak to be reliably resolved here in this representation.
The triangle mode rapidly loses spectral weight with increasing magnetic field and is no longer resolved above approximately \SI{1}{T}.

For \mbox{$h^\omega \parallel b$} [see Fig.~\ref{fig:FTIR_Hpa}(b)(c)], the spectra exhibit richer behaviors.
Already in zero magnetic field, the previously identified low-energy magnon mode associated predominantly with the ferromagnetic Cu1 chains, marked by asterisks, gives rise to a strong absorption feature near \SI{1.2}{meV}.
With increasing $B$, this mode shifts moderately to higher energies and disappears above \SI{4.75}{T}.

At higher energies several modes are observed between \SI{3.9}{meV} and \SI{5}{meV}, marked by triangles, crosses, circles, and plus symbols, consistent with the zero-field data shown in Fig.~\ref{fig:FTIR_Tdep}.
This energy range coincides with the regime in which the spinon continuum associated with the antiferromagnetic Cu2 chains overlaps and interacts with magnon excitations originating from the ferromagnetic Cu1 chains, as previously revealed by inelastic neutron scattering~\cite{Zhang_2020}.
%Related high-frequency excitations have been interpreted as spinon--magnon bound states in previous terahertz spectroscopy measurements~\cite{Povarov_2024}.

Above about \SI{4.75}{T}, all these low-field excitations are suppressed and cannot be resolved, indicating a field-induced reconstruction of the magnetic excitation spectrum.
This characteristic field scale is associated with the spin-flop transition for \mbox{$B\parallel a$}, seen in the magnetization data in Figs.~\ref{fig:Magnetization_abc}(a) and \ref{fig:Magnetization_abc}(d)~\cite{Xiao_2022, Povarov_2024, Zhao_2019}.
%Small differences in the extracted transition field may arise from sample-to-sample variations and from the smoothing applied to the magnetization data.
At the spin-flop transition, the spins of the ferromagnetic Cu1 chains reorient from their zero-field direction close to the diagonal of the $ac^*$ into the $ab$ plane, where spins of the ferromagnetic chains point alternately along the $+b$ and $-b$ directions, as illustrated schematically in Fig.~\ref{fig:crystalstructure}(c).
The polarization of the terahertz magnetic field along the same direction (i.e. $h^\omega \parallel b$) is hence not able to excite those otherwise active magnetic modes.

In contrast, above $B_{\mathrm{SF},a}$ a distinctly different high-field excitation spectrum is resolved.
The most prominent feature is the $\beta$ mode, which appears immediately above the transition and is resolved as two closely spaced branches.
With increasing magnetic field, both $\beta$ branches soften and gradually lose spectral weight.
Moreover, a weaker low-energy $\gamma$ mode appears at high fields near the lower edge of the spectral window and hardens with increasing field.
The observation of the field-split $\beta$ mode and the gradually enhanced low-energy $\gamma$ mode reflects a substantial field-induced reconstruction of the magnetic structure.
Additional weaker modes were resolved by field-swept electron spin resonance measurements using narrowband sources \cite{Povarov_2024}, which also confirms a field-induced spin reorientation \cite{Xiao_2022,Povarov_2024}.

%-------------------------------------------------------------------------------------------------------------------------

\begin{figure}[h!]
    \centering
    \includegraphics[width=1\linewidth]{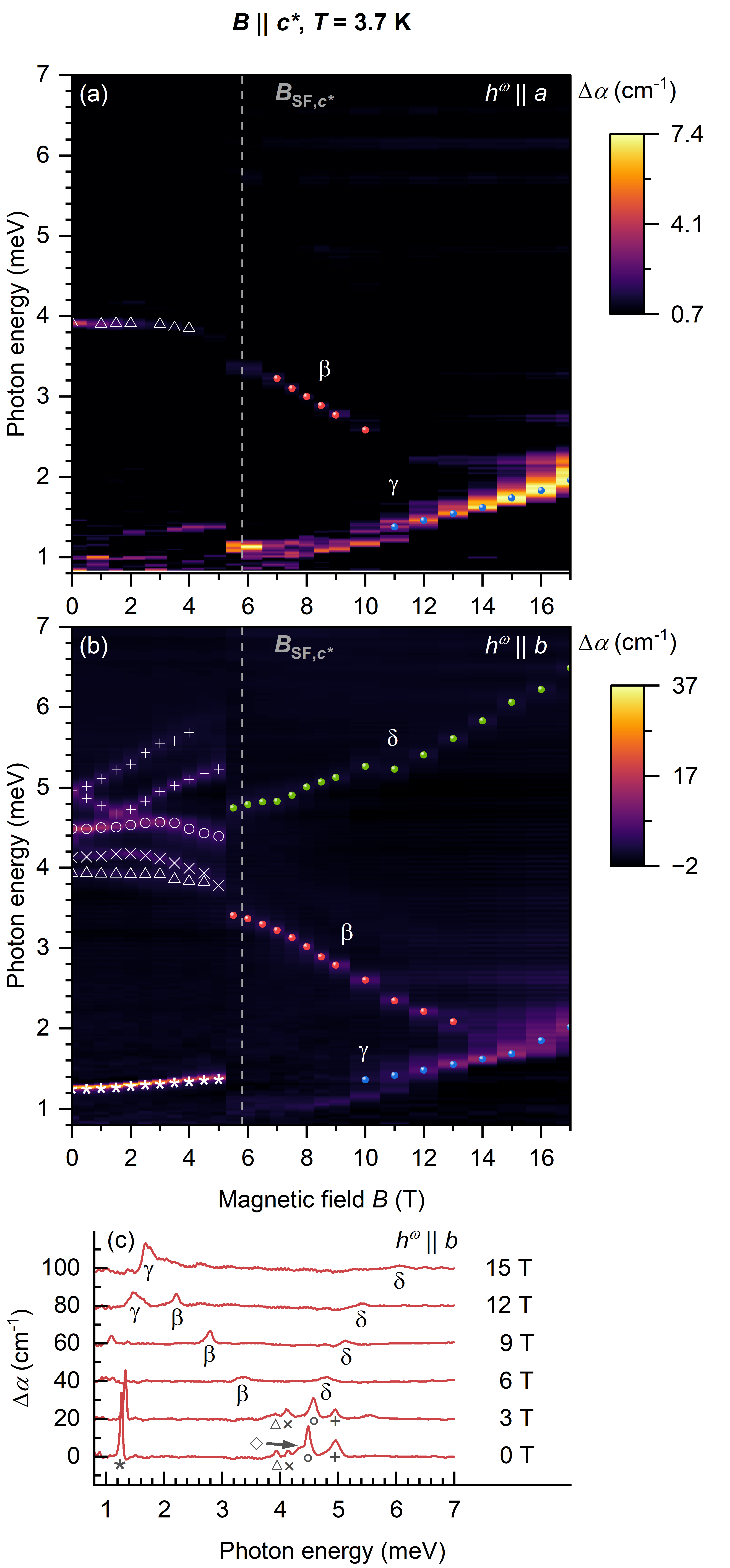}
    \caption{Magnetic-field dependence of the differential absorption spectrum $\Delta\alpha$ of \ce{Cu2(OH)3Br} at \SI{3.7}{K} for $B\parallel c^*$, where (a) $h^\omega \parallel a$ and (b) $h^\omega \parallel b$.
    The spin-flop transition at $B_{\mathrm{SF},c^*}$ is highlighted by the grey dashed line. (c) Selected spectra for \mbox{$h^\omega \parallel b$} at different magnetic fields. The spectra are offset for clarity.}
    \label{fig:FTIR_Hpc}
\end{figure}

\subsubsection{\label{subsec:FTIR_Bdep_c}Field dependence for $B \parallel c^*$}

For $B \parallel c^*$ the differential absorption spectra are shown in Fig.~\ref{fig:FTIR_Hpc}, while the extracted absorption maxima are summarized in Fig.~\ref{fig:exmodes_maxima}(b).
The overall field-dependent evolution of the excitation spectra resembles that observed for \mbox{$B\parallel a$}, while the absorption lines more intense, persist over a broader field range.

For the polarization \mbox{$h^\omega \parallel a$} [Fig.~\ref{fig:FTIR_Hpc}(a)], both the triangle and the lower branch of the $\beta$ mode are clearly resolved.
Above approximately \SI{10}{T}, an additional broad low-energy feature, denoted as the $\gamma$ mode, becomes visible at the lower boundary of the measured spectral range.
With increasing $B$, this feature shifts to higher energies toward the $\beta$ mode.
At \SI{17}{T}, they form a rather broad absorption feature centered near \SI{2}{meV}.

\begin{figure}[t]
    \centering
    \includegraphics[width=1\linewidth]{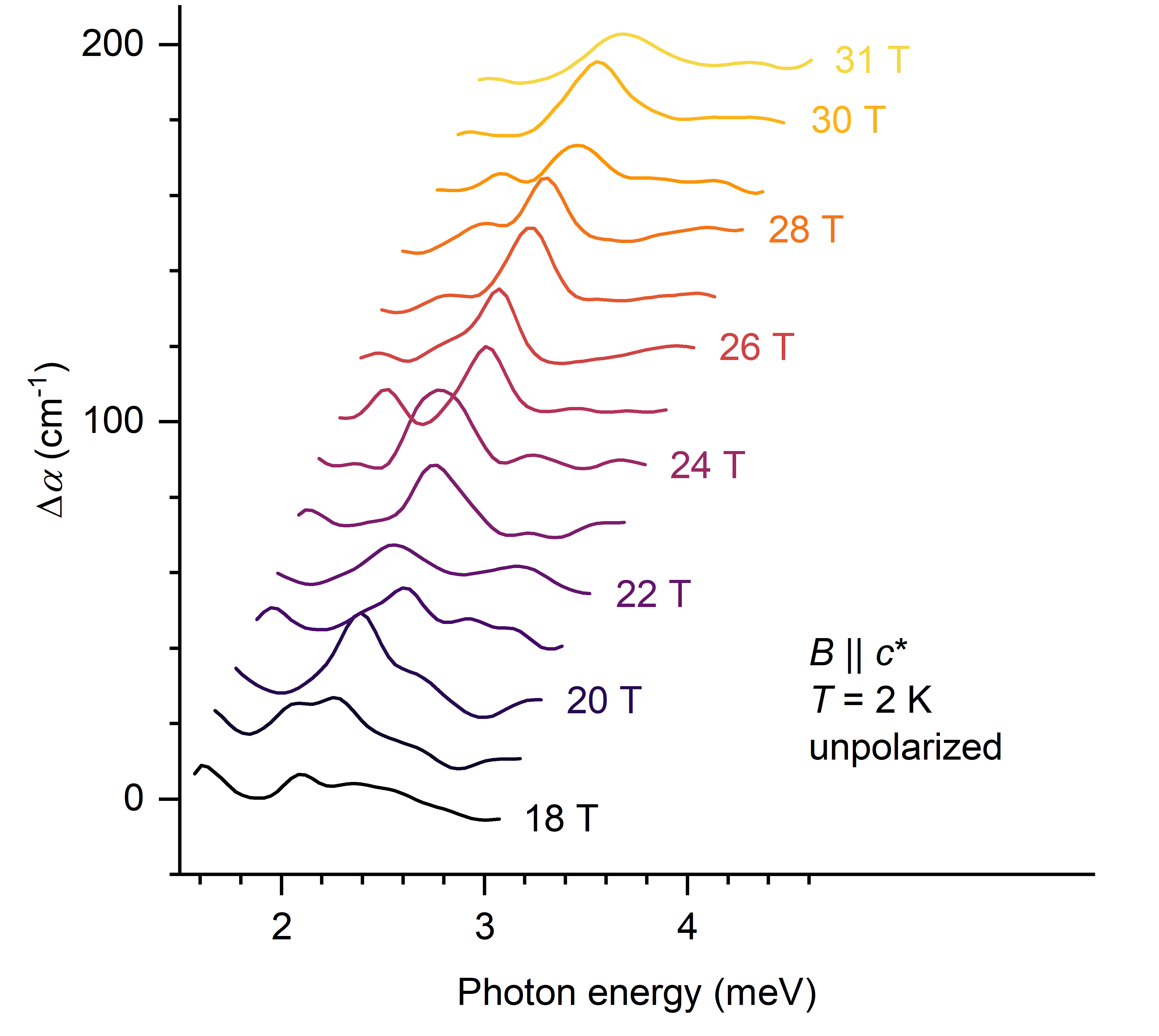}
    \caption{Unpolarized differential absorption spectra $\Delta\alpha$ of \ce{Cu2(OH)3Br} at \SI{2}{K} for $B\parallel c^*$ up to \SI{31}{T}.}
    \label{fig:FTIR_Hpc_Nijmegen}
\end{figure}

For the polarization \mbox{$h^\omega \parallel b$} [see Fig.~\ref{fig:FTIR_Hpc}(b)(c)], the low-energy magnon mode, marked by asterisks, hardens only weakly with increasing field.
Compared with \mbox{$B\parallel a$}, the triangle and cross modes remain visible up to higher fields.
While the triangle mode exhibits only a weak field dependence, the cross mode softens nonlinearly and approaches the triangle mode.
The circle mode is nearly field independent at low fields but develops a kink-like field dependence near \SI{3.5}{T}, above which this mode softens.

Already above about \SI{5}{T}, all the observed low-field modes disappear within a narrow field interval.
This abrupt change coincides with the spin-flop transition observed in the magnetization for \mbox{$B\parallel c^*$} [Fig.~\ref{fig:Magnetization_abc}(b)(e)] and reported previously~\cite{Xiao_2022,Povarov_2024}.
The resulting spin configuration is analogous to that illustrated schematically for \mbox{$B\parallel a$} in Fig.~\ref{fig:crystalstructure}(c), with the Cu1 moments reoriented predominantly along the $b$ direction but slightly canted toward the applied field along $c^*$.
We therefore associate the reconstruction of the excitation spectrum with the field-induced spin-flop transition at $B_{\mathrm{SF},c^*}$.

Above the spin-flop transition, three distinct excitations are resolved.
The $\beta$ mode persists as a single softening branch, in contrast to the pair of closely spaced branches observed for \mbox{$B\parallel a$}.
At lower energies, the $\gamma$ mode hardens with increasing field.
In addition, a distinct higher-energy $\delta$ mode emerges near \SI{4.7}{meV} immediately above the spin-flop transition and shifts towards higher energies in high fields, reaching approximately \SI{6.6}{meV} at \SI{17}{T}.

To extend the accessible field range beyond \SI{17}{T}, we performed unpolarized differential absorption measurements for $B \parallel c^*$ up to \SI{31}{T}, as shown in Fig.~\ref{fig:FTIR_Hpc_Nijmegen}.
To obtain field-dependent features, we calculate the differential absorption spectra by using the average of all measured field-dependent transmission spectra as a reference.
The low-energy $\gamma$ branch can be continuously tracked into the high-field regime as a broad absorption feature up to the highest field [see also Fig.~\ref{fig:exmodes_maxima}(b)].
The field dependence of the low-energy branch was fitted using a linear field dependence according to the Zeeman interaction, which yields a Landé factor of $g\approx1.97\pm0.03$, as indicated by the purple dashed line in Fig.~\ref{fig:exmodes_maxima}(b).
Field-induced spin-flop transition was also confirmed by field-swept electron spin resonance measurements using narrowband sources~\cite{Xiao_2022,Povarov_2024}.

%-------------------------------------------------------------------------------------------------------------------------

\subsubsection{\label{subsec:FTIR_Bdep_b}Field dependence for \mbox{$B\parallel b$}}

\begin{figure}[t]
    \centering
    \includegraphics[width=1\linewidth]{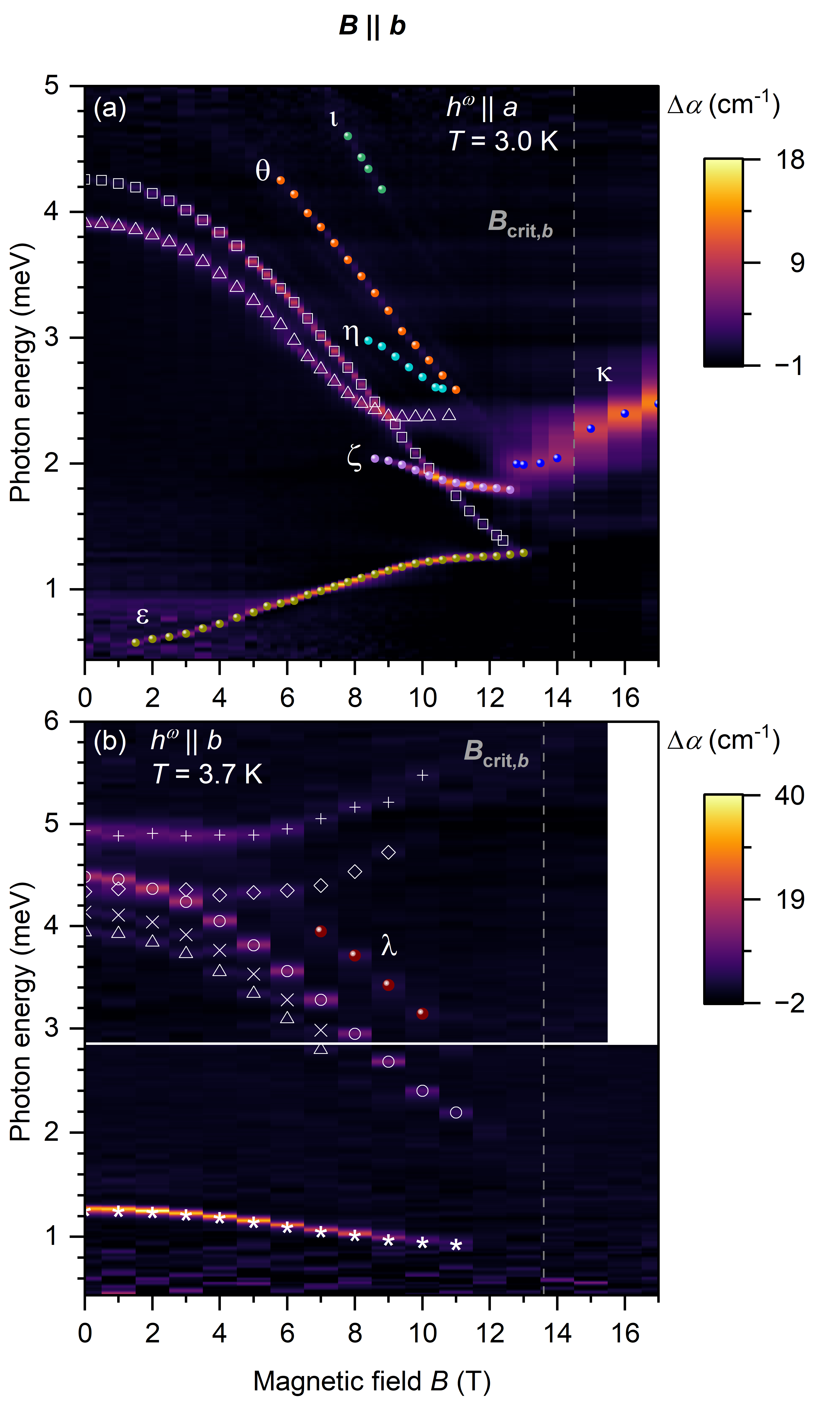}
    \caption{Magnetic-field dependence of the differential absorption spectrum $\Delta\alpha$ of \ce{Cu2(OH)3Br} for an applied magnetic field \mbox{$B\parallel b$} with terahertz polarizations (a) 
    \mbox{$h^\omega \parallel a$} at \SI{3}{K} and (b) \mbox{$h^\omega \parallel b$} at \SI{3.7}{K}, respectively.
    For \mbox{$h^\omega \parallel b$}, data from two spectral filters were combined, as indicated by the white line. The critical fields $B_{\mathrm{crit},b}$ obtained from magnetization measurements are marked by grey dashed lines.}
    \label{fig:FTIR_Hpb}
\end{figure}

\begin{figure*}[t]
    \centering
    \includegraphics[width=1\linewidth]{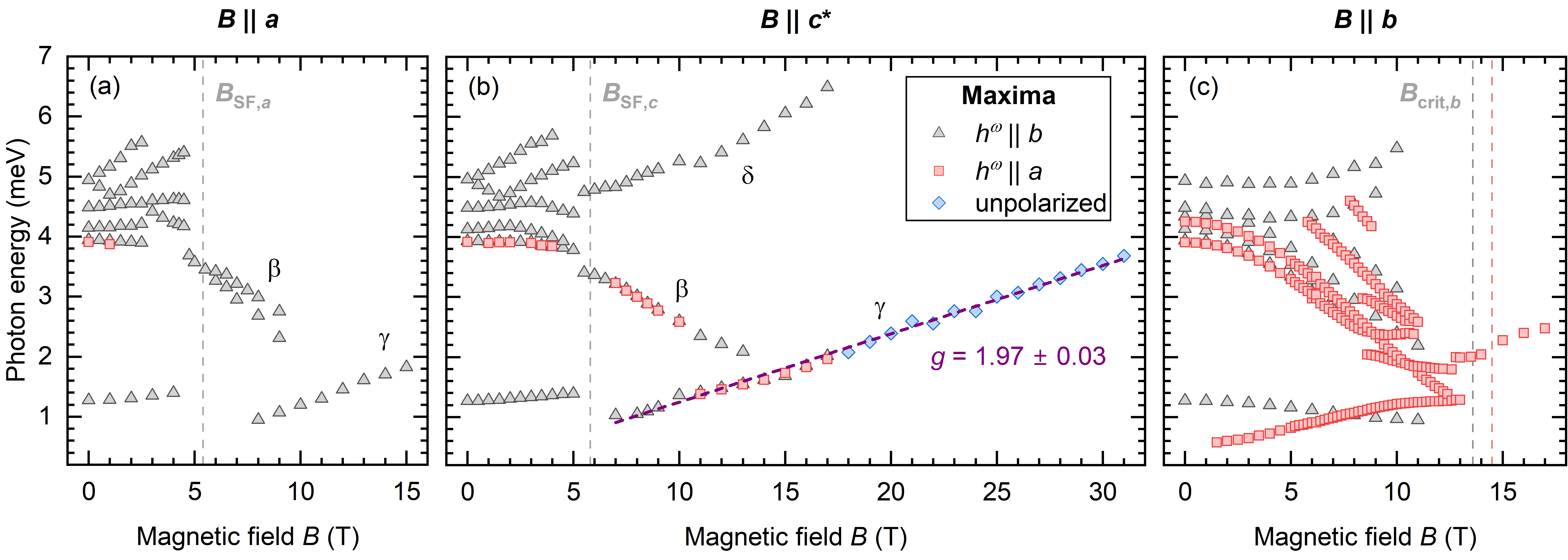}
    \caption{Extracted absorption maxima from the spectra shown in Figs.~\ref{fig:FTIR_Hpa}, \ref{fig:FTIR_Hpc}, \ref{fig:FTIR_Hpc_Nijmegen}, and \ref{fig:FTIR_Hpb} for (a) \mbox{$B\parallel a$}, (b) \mbox{$B\parallel c^*$}, and (c) \mbox{$B\parallel b$}.
    The spin-flop transitions at $B_{\mathrm{SF},a}$ and $B_{\mathrm{SF},c^*}$ as well as the critical field $B_{\mathrm{crit},b}$ are indicated by dashed lines.
    The purple dashed line in (b) represents a linear fit used to extract the effective $g$ factor.}
    \label{fig:exmodes_maxima}
\end{figure*}

The absorption spectra for \mbox{$B\parallel b$} are shown in Fig.~\ref{fig:FTIR_Hpb} and the eigenenergies of the observed modes as a function of the applied field for two different terahertz polarizations are summarized in Fig.~\ref{fig:exmodes_maxima}(c).
In contrast to the other two applied field orientations, \mbox{$B\parallel b$} is transverse to the ordered spins in both antiferromagnetic and ferromagnetic chains [see Fig.~\ref{fig:crystalstructure}(a)].
Instead of a spin-flop transition, a continuous magnetization increase is observed [Fig.~\ref{fig:Magnetization_abc}(c)], characterized by the gradual polarization of the Cu1 spin chains along the field direction [Fig.~\ref{fig:crystalstructure}(d)] \cite{Reinold_2025}.

%corresponding to a terahertz magnetic field transverse to the spin chains.
%This polarization channel is particularly interesting in view of the proposed connection between the field-induced transition in \ce{Cu2(OH)3Br} and transverse-field Ising-chain quantum criticality~\cite{Reinold_arXiv_2026}.
%In quasi-one-dimensional systems such as \ce{CoNb2O6}~\cite{Coldea_2010,Amelin_2020,Amelin_2022} and \ce{BaCo2V2O8}~\cite{Amelin_2022,Zhang_E8_2020,Zou_2021}, an emergent $E_8$ excitation spectrum has been observed in the ordered phase close to such a field-induced transition.

For the terahertz polarization \mbox{$h^\omega\parallel a$}, the triangle mode evolves continuously from its zero-field position near \SI{3.9}{meV} and exhibits a pronounced nonlinear field-dependent softening with increasing field.
A second branch, marked by the square symbols, extrapolates to approximately \SI{4.3}{meV} at zero field [see also Fig.~\ref{fig:FTIR_Tdep}(a)].
Although its spectral weight is very small at zero field, it becomes increasingly pronounced upon application of a magnetic field, while simultaneously exhibiting rapid softening.
With increasing field, the two branches approach each other and cross at approximately \SI{9}{T} and \SI{2.4}{meV}.
No discernible mode repulsion or avoided crossing is observed at their intersection, indicating that the two excitations retain their distinct character and hybridize only weakly, if at all.
This is consistent with our assignment of the square mode as being excitations of the antiferromagnetic Cu2 chains and of the triangle mode as a magnon mode of the ferromagnetic Cu1 chains, according to the spin-wave theory simulation and inelastic neutron scattering measurements at zero field \cite{Zhang_2020}.
Beyond the crossing, the square branch continues to soften, reaching approximately \SI{1.4}{meV} at \SI{12.4}{T}, whereas the triangle branch can be traced to approximately \SI{10.8}{T} at \SI{2.6}{meV}, above which it broadens substantially.

At lower energies, a qualitatively different field dependence is observed.
A further excitation branch, labeled $\epsilon$, first becomes resolvable at a finite field of \SI{1.5}{T} near \SI{0.6}{meV}.
In contrast to the softening branches discussed above, this excitation hardens nonlinearly with increasing field and reaches approximately \SI{1.3}{meV} close to \SI{13}{T}.
Since this mode is well below the energy of the spinon continuum in the antiferromagnetic chains, it is natural to assign the $\epsilon$ mode as a magnon in the ferromagnetic chains.

Upon further field increase, several additional excitation branches become resolved in an intermediate field range.
A branch labeled $\zeta$ appears at approximately \SI{8.6}{T} near \SI{2}{meV} and exhibits a weak nonlinear softening with increasing field.
It can be tracked down to approximately \SI{1.8}{meV} at \SI{12.6}{T}, above which it is no longer distinguishable as an individual excitation.
At higher energies, a weak branch labeled $\eta$ becomes visible around \SI{8.4}{T} near \SI{3}{meV} and softens to approximately \SI{2.6}{meV} at \SI{10.6}{T}.

A further weak excitation, labeled $\theta$, becomes resolvable already at approximately \SI{5.8}{T} near \SI{4.3}{meV} and exhibits a considerably stronger nonlinear softening, reaching a similar energy of approximately \SI{2.6}{meV} at \SI{11}{T}.
Together with the triangle branch, the $\eta$ and $\theta$ modes therefore converge toward a common energy range around \SIrange{10}{11}{T}.
At the same time, their individual spectral signatures progressively lose spectral weight and can be no longer clearly separated, resulting in a broad and weakly resolved response.

A considerably weaker high-energy excitation, labeled $\iota$, is resolved only over a narrow field interval.
It softens approximately linearly from about \SI{4.6}{meV} at \SI{7.8}{T} to \SI{4.2}{meV} at \SI{8.8}{T}.
This mode with a much higher energy than the magnon modes is possibly an excitation of antiferromagnetic chains.

The appearance of the modes (especially $\epsilon$, $\zeta$, triangle, $\theta$) in the intermediate field range below the critical field could be due to the emergence of $E_8$ dynamics (where $E_8$ is an exceptional Lie algebra) \cite{Zamolodchikov89,Coldea_2010,Amelin_2020,Amelin_2022,Zhang_E8_2020,Zou_2021} in the ferromagnetic chains \cite{Reinold_arXiv_2026}, which is driven close to the transverse-field Ising-chain-type quantum phase transition and perturbed by subleading interchain interactions.  
A detailed discussion of the emergent energy scale due to an integable field theory close to the transverse-field Ising-chain-type quantum phase transition can be found in Ref.~\cite{Reinold_arXiv_2026}.

The quantum spin dynamic response exhibits a sharp change when crossing the critical field \mbox{$B_{\mathrm{crit},b}$}.
In contrast to several modes below \mbox{$B_{\mathrm{crit},b}$}, one broad and intense mode (labeled as $\kappa$) is present above \mbox{$B_{\mathrm{crit},b}$}.
With increasing field, the $\kappa$ mode hardens continuously to approximately \SI{2.5}{meV} at the highest field of \SI{17}{T}, while gaining spectral weight and becoming progressively better defined.
The evolution of the excitation spectrum around \mbox{$B_{\mathrm{crit},b}$} is consistent with the occurrence of field-induced effective decoupling of the two types of chains and a dimensional reduction at \mbox{$B_{\mathrm{crit},b}$}, which was established from magnetization measurements and quantum Monte Carlo calculations~\cite{Reinold_2025,Khatua_arXiv_2026}.

For \mbox{$h^\omega\parallel b$}, corresponding to a terahertz magnetic field parallel to both the applied static field and the spin-chain direction, a qualitatively different spectral evolution is observed.
In contrast to the transverse polarization channel, the response is dominated by the continuous evolution of excitations already present at zero field, with only one additional weak branch (labeled as $\lambda$) becoming resolved at higher fields.

The lowest-energy excitation, marked by the asterisk symbols and previously associated with magnonic character, exhibits only a weak softening from approximately \SI{1.3}{meV} at zero field to about \SI{1}{meV} at \SI{11}{T}.
At higher energies, the triangle, cross, and circle branches display a similar pronounced nonlinear softening with increasing field, comparable to their evolution for the other field orientations.
The triangle mode decreases from approximately \SI{4}{meV} at zero field to \SI{2.8}{meV} at about \SI{7}{T}, while the closely lying cross mode softens from approximately \SI{4.1}{meV} to \SI{3}{meV} over a similar field range.
The circle mode exhibits the strongest softening within this group and remains resolvable to considerably higher fields, evolving from approximately \SI{4.5}{meV} at zero field to about \SI{2.2}{meV} at \SI{11}{T}.
While the triangle and cross branches lose spectral weight and become unresolved above approximately \SI{7}{T}, the circle mode can therefore be followed almost up to the critical-field $B_{\mathrm{crit},b}$.

In contrast to these softening excitations, the diamond and plus branches shift toward higher energies with increasing field.
The diamond mode hardens nonlinearly from approximately \SI{4.4}{meV} at zero field to about \SI{4.8}{meV} at \SI{9}{T}.
During this evolution, it crosses the strongly softening circle branch near \SI{2}{T}.
No discernible mode repulsion or avoided crossing is observed at their intersection, suggesting that the two excitations retain their distinct character and exhibit at most weak hybridization.
The plus mode follows a similar hardening trend but remains considerably more intense, increasing from approximately \SI{4.9}{meV} at zero field to about \SI{5.5}{meV} at \SI{10}{T}.

At intermediate fields, an additional weak excitation branch, labeled $\lambda$, becomes resolvable near \SI{3.9}{meV} at approximately \SI{7}{T}.
It subsequently softens to about \SI{3.1}{meV} at \SI{10}{T}, approximately following the field-dependent evolution of the triangle, cross, and circle branches.
Above approximately \SI{11}{T}, all the excitations rapidly lose spectral weight and become unresolvable.

The disappearance of the magnetic absorption modes observed for \mbox{$h^\omega\parallel b$} can be understood as follows.
On the one hand, above \mbox{$B_{\mathrm{crit},b}$}, the Cu1 spins are aligned by the magnetic field along the $b$ axis, which is parallel to the THz magnetic field.
Since the THz magnetic field \mbox{$h^\omega\parallel b$} is parallel to the polarized Cu1 moments, it couples only weakly to their transverse spin excitations, rendering these modes inactive in this polarization channel.
On the other hand, the field-induced dimensional reduction leads to the suppression of the three-dimensional magnetic order, and enhances one-dimensional fluctuations in the antiferromagnetic chain.
The characteristic spinon excitations of the antiferromagnetic chains should form a continuum, which is not captured in this transverse channel, but can be much better resolved by inelastic neutron scattering experiment.

%-------------------------------------------------------------------------------------------------------------------------

\begin{figure}[t]
    \centering
    \includegraphics[width=1.0\linewidth]{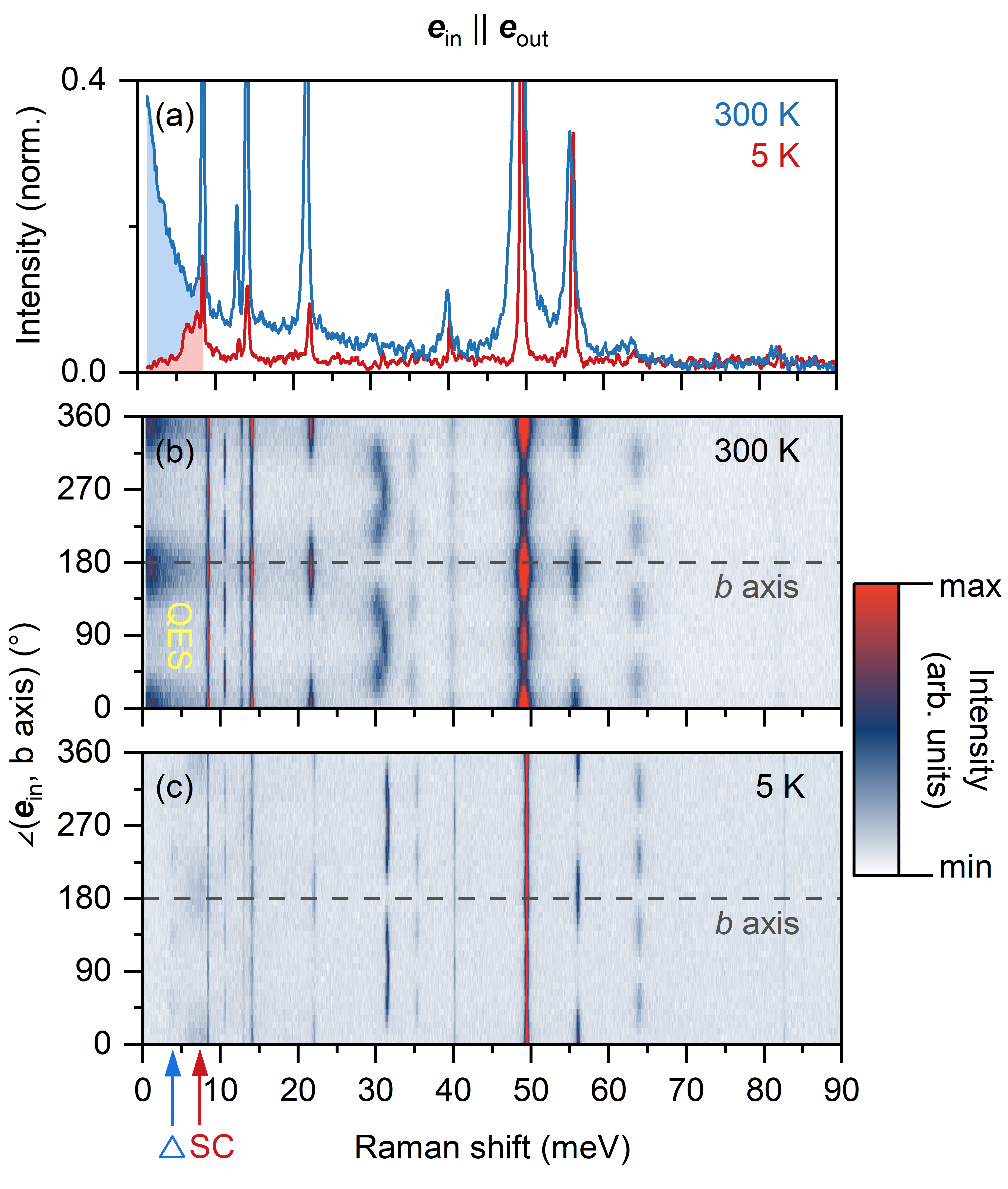}
    \caption{Raman scattering data measured in parallel polarization $(\bm{e}^\mathrm{in}\parallel \bm{e}^\mathrm{out})$.
    (a) Raman scattering spectra at $\SI{300}{K}$ and $\SI{5}{K}$, normalized for clarity.
    (b)(c) Color-contour plots of the Raman scattering intensity as a function of in-plane light polarization at $\SI{300}{K}$ and $\SI{5}{K}$, respectively.
    Grey dashed lines mark the spin-chain direction ($\parallel b$). The magnon excitation (triangle), spinon continuum (SC), and quasielastic scattering (QES) are indicated.}
    \label{fig:Raman_PolDep}
\end{figure}

\subsection{\label{subsec:Raman}Raman spectroscopy}

\subsubsection{Polarization and temperature dependence at $B=0$}

To further investigate the polarization dependence and thermal evolution of the excitations across $T_\mathrm{N}$, as well as the interplay among spinons, magnons, and phonons, we now turn to Raman spectroscopy.

In Fig.~\ref{fig:Raman_PolDep}(a) two Raman spectra taken in parallel configuration ($\bm{e}_\mathrm{in}\parallel \bm{e}_\mathrm{out}$) at \SI{300}{K} and at \SI{5}{K} are shown.
At high temperatures, the low-energy spectrum below $\SI{10}{meV}$ is dominated by significant quasi-elastic scattering, which is suppressed in the \SI{5}{K} spectrum, with the Raman scattering intensity approaching 0 as the Raman shift approaches zero.

In Fig.~\ref{fig:Raman_PolDep}(b) and \ref{fig:Raman_PolDep}(c) we show detailed polarization plots as a function of $e_\text{in}$ taken at \SI{300}{K} and at \SI{5}{K}, respectively.
These plots reveal selection rules for individual excitations. We note that the quasi-elastic signal in panel (b), labeled QES, has a clear 2-fold symmetry, appearing with a periodicity of 180$^{\circ}$.
Pronounced quasi-elastic scattering is a hallmark of low-dimensional quantum magnets.
It arises from thermally activated spin fluctuations in the absence of long-range order, and in quasi-1D spin chain systems it can be probed when the light polarization is parallel to the chain direction~\cite{Reiter_1976, Lemmens_2003}.
We can therefore identify this feature as a fingerprint of thermally activated fluctuating spin degrees of freedom and assign the chain direction along its maximum intensity.
Cooling the sample below $T_\mathrm{N}$ quenches the quasi-elastic scattering related to thermally-induced spin fluctuations and opens an excitation gap of $\sim \SI{1.2}{meV}$ in the spin excitation spectrum~\cite{Zhang_2020}.
Above this gap, distinct excitations appear: a sharp excitation at about \SI{4}{meV} with a four-fold periodicity [see Figure~\ref{fig:Raman_PolDep}(c)], and broader excitations centered between \SI{5}{meV} and \SI{10}{meV} with a two-fold periodicity.
The former one is identified as the magnon marked by a triangle. Its four-fold periodicity is also consistent with the fact that quenches the quasi-elastic scattering related to thermally-induced spin fluctuationse can probe it through terahertz spectroscopy in both \mbox{$h^\omega \parallel a$} and \mbox{$h^\omega \parallel b$}.
On the other hand, the direction of the two-fold periodicity for broader excitations coincides with that of the high-temperature quasi-elastic signal.
Their distinct polarization dependences indicate that the excitations at \SI{4}{meV} and those between 5-10\,meV are different in nature.
To substantiate this assumption, we turn to their temperature- and field-dependent evolution.

\begin{figure}[t]
    \centering
    \includegraphics[width=1\linewidth]{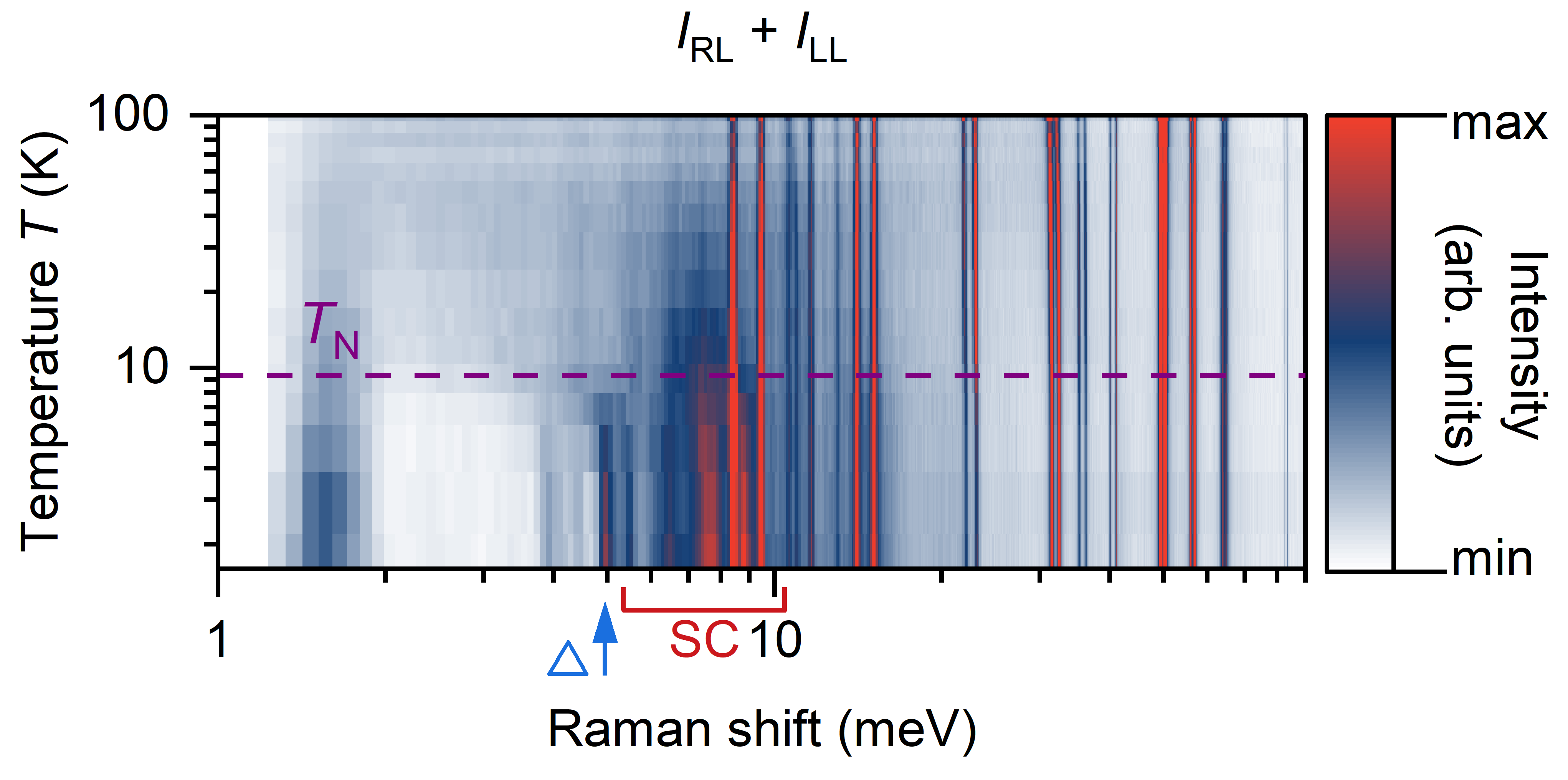}
    \caption{Color-contour plot of the Bose-corrected Raman scattering intensity $I_{\mathrm{RL}} + I_{\mathrm{LL}}$ measured between \SI{1.6}{K} and \SI{100}{K}, obtained by summing the cross-circular $I_{\mathrm{RL}}$ and co-circular $I_{\mathrm{LL}}$ polarization channels.
    The N\'eel temperature $T_\mathrm{N}=\SI{9.3}{K}$ is indicated by the purple dashed line. The magnon excitation (triangle) is marked by the blue arrow, while the spinon continuum (SC) is highlighted by the red bracket.}
    \label{fig:Raman_Tdep_BOSE}
\end{figure}

In Fig.~\ref{fig:Raman_Tdep_BOSE} we show color contour plots of the Bose-corrected Raman scattering intensity as a function of temperature in unpolarized configuration.
To emphasize the low-energy excitations, we employ a log-log scale for energy and temperature.
In this dataset we can directly trace the fate of the sharp excitation around \SI{4}{meV}, marked by a triangle: as the temperature is raised beyond the magnetic ordering temperature $T_\mathrm{N}$, the excitation abruptly vanishes.
Such thermal behavior, together with the narrow linewidth and overall weak scattering intensity, is consistent with a one-magnon excitation.
Next we turn our attention to the broader excitations inhabiting the energy range 5-10 meV.
These features exhibit a distinct change across $T_\mathrm{N}$, suggesting spinon-magnon hybridization or the confinement of spinons.
However, in contrast to one-magnon modes, a significant scattering intensity remains above $T_\mathrm{N}$ within the same energy range that only gradually diminishes and re-normalizes toward a broader quasi-elastic contribution at around $T = \SI{100}{K}$.
This evolution with temperature is uncharacteristic for one-magnon excitations, and instead follows that expected for spinon excitations~\cite{Wulferding_2020, Wulferding_2025}.
Consistent with this picture, the two-fold periodicity of these excitations signifies that they are intimately linked to the spin chains. 
A spinon continuum constitutes the fundamental excitations of (quasi-) 1D spin chains, which describes the magnetic subsystem of Cu$_2$(OH)$_3$Br above $T_\mathrm{N}$.
The structured excitation spectrum with relatively sharp modes that emerges our of the featureless spinon continuum below $T_\mathrm{N}$ suggests the possibility of spinon-magnon hybridization or the formation of confined spinons that coexist with lower-energy magnons and are stabilized by the 2D magnetic exchange topology.

\begin{figure}[t]
    \centering
    \includegraphics[width=1.0\linewidth]{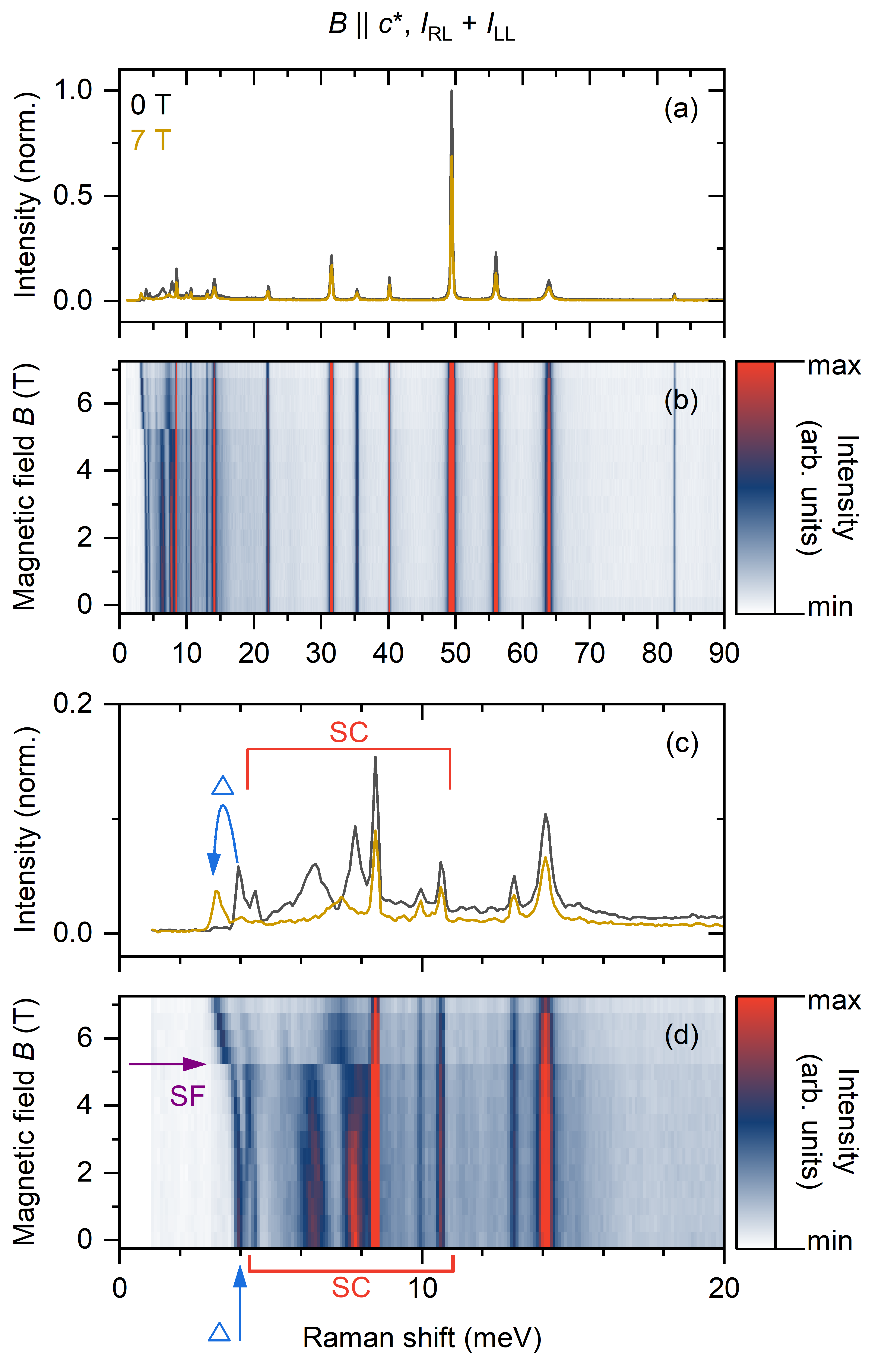}
    \caption{Magnetic-field-dependent Raman scattering spectra measured up to \SI{7}{T} for $B\parallel c^*$ at $T=\SI{1.6}{K}$. The displayed intensity corresponds to the sum of the cross-circular and co-circular polarization channels, $I_{\mathrm{RL}}+I_{\mathrm{LL}}$.
    (a) Raman scattering spectra up to a Raman shift of \SI{90}{meV} at $B=\SI{0}{T}$ (black) and \SI{7}{T} (yellow), normalized for clarity.
    (b) Color-contour plot of the Raman scattering intensity as a function of magnetic field up to a Raman shift of \SI{90}{meV}.
    (c) Low-energy Raman spectra corresponding to the data shown in (a) and (b), highlighting the magnetic excitation spectrum. The magnon excitation (triangle) and spinon continuum (SC) are marked accordingly.
    (d) Corresponding low-energy color-contour plot. The spin-flop (SF) transition at $B_{\mathrm{SF}}\approx\SI{5}{T}$ is indicated by the purple arrow.}
    \label{fig:Raman_BBdep}
\end{figure}

\subsubsection{Field dependence for $B \parallel c^*$}

Finally, we comment on the fate of magnetic excitations across the spin-flop transition in applied fields along the crystallographic $c^*$ axis.
Figure~\ref{fig:Raman_BBdep}(a) shows two individual Raman spectra taken at the base temperature ($T = \SI{1.6}{K}$) in $B = \SI{0}{T}$ and $B = \SI{7}{T}$.
While Raman excitations above \SI{10}{meV} (corresponding to phonons) are largely unaffected by the magnetic field, the low-energy range, dominated by magnetic excitations, undergoes some changes.
The evolution of magnetic and phononic Raman-active modes is displayed in the color-contour plot of Fig.~\ref{fig:Raman_BBdep}(b), confirming the rather static phonon behavior, while abrupt changes in the magnetic excitation spectrum occur at the spin-flop transition around \SI{5}{T}.
In Figs.~\ref{fig:Raman_BBdep}(c) and \ref{fig:Raman_BBdep}(d) we focus on these excitations by zooming into the low-energy range below \SI{20}{meV}, to further highlight the abrupt nature of the transition.
The spin-flop transition reorients the magnetic sublattices, modifying the effective coupling between magnetic excitations by partially decoupling the ferromagnetic and antiferromagnetic sublattices, thereby reducing the number of distinct magnon modes observed.
Particularly, the sharp modes due to interactions of the ferromagnetic and antiferromagnetic chains are replaced by a nearly symmetric, slightly broad continuum.
We summarize the magnetic excitation spectrum in Cu$_2$(OH)$_3$Br as follows: below $T_\mathrm{N}$ and $B_{\mathrm{SF}}$ magnons coexist and interact with spinon excitations.
As the magnetic field increases above $B_{\mathrm{SF}}$ we observe a coexistence between a reduced magnon spectrum and enhanced spinon continuum.
Meanwhile, for temperatures above $T_\mathrm{N}$ all magnons have melted, leaving only a continuum of entangled spinons that gradually crosses over into thermally-induced, disentangled spin fluctuations at around \SI{100}{K}.

\section{\label{sec:Conclusion}Conclusion}

To summarize, our results establish a strongly anisotropic magnetic field-dependent evolution of the magnetic excitation spectrum of \ce{Cu2(OH)3Br}, governed by the interactions within the ferromagnetic Cu1 and the antiferromagnetic Cu2 spin-chain subsystems as well as the interchain interactions.
While the applied fields along $a$ and $c^*$ induce spin-flop transitions, the field applied along the chain direction $b$ (i.e. transverse to the ordered spins) drives a continuous evolution toward the polarization of the Cu1 subsystem, as revealed by our magnetization measurements.

By performing polarization-resolved terahertz and Raman spectroscopic measurements, we have resolved and traced the magnetic excitations for different field orientations and over a broad magnetic-field range, corresponding to different sectors of dynamical spin correlations.
We observed sharp contrast in the characteristic magnetic dynamics crossing the field-induced phase transitions.
Tracking the individual excitation branches in the applied fields further allows us to distinguish modes corresponding to the spin dynamics in the ferromagnetic or antiferromagnetic chains.
In particular, the observed field dependencies reveal how the overlapping magnon and spinon sectors progressively reorganize as the balance between the two magnetic subsystems is modified.

At high fields above the field-induced phase transitions, we resolved less excitation modes than below, and observed broadening of the excitations that is consistent with the weakening effective coupling between the ferromagnetic and antiferromagnetic chains and enhanced one-dimensional spin fluctuations.
Our results therefore provide a unified spectroscopic picture of how collective and fractionalized spin excitations evolve with magnetic field crossing field-induced phase transitions in a coupled ferromagnetic--antiferromagnetic spin-chain system.
These findings strongly motivate inelastic neutron scattering studies of the interplay of magnon and spinon dynamics in the applied magnetic fields and in the disordered phases with enhanced low-dimensional quantum fluctuations. 

\begin{acknowledgments}
We acknowledge helpful discussions with F. F. Assaad and M. Raczkowski. This work was partially supported by the European Research Council (ERC) under the Horizon 2020 research and innovation programme, Grant Agreement No.~950560 (DynaQuanta).
D.W. was supported by the Institute of Information \& Communications Technology Planning \& Evaluation (IITP)-ITRC (Information Technology Research Center) grant funded by the Korea government (MSIT) (IITP-2026-RS-2024-00437191) and GRDC Cooperative Hub Program funded by the Korea government (NRF) (RS-2023-00258359).
The work conducted in Tallinn was supported by the Estonian Research Council under Grant No.~PRG736.
We acknowledge funding from the Deutsche Forschungsgemeinschaft (DFG, German Research Foundation) under Project-ID 277146847—CRC 1238 (subproject B01).
We also acknowledge the support of the HLD at HZDR, member of the European Magnetic Field Laboratory (EMFL) and of HFML-FELIX, member of the European Magnetic Field Laboratory (EMFL).
\end{acknowledgments}

\bibliographystyle{apsrev4-2}
\bibliography{COHB_bib}% Produces the bibliography via BibTeX.

\end{document}